\documentclass[10pt,journal,compsoc,final]{IEEEtran}

\ifCLASSOPTIONcompsoc
  \usepackage[nocompress]{cite}
\else
  \usepackage{cite}
\fi

\ifCLASSINFOpdf
  \usepackage[pdftex]{graphicx}
  \graphicspath{figures}
  \DeclareGraphicsExtensions{.pdf,.eps,.png, .eps}
\else
  \usepackage[dvips]{graphicx}
  \graphicspath{figures}
  \DeclareGraphicsExtensions{.pdf,.eps,.png, .eps}
\fi
\graphicspath{{./figures/}}

\usepackage{amsmath}

\usepackage{array} 

\ifCLASSOPTIONcompsoc
  \usepackage[caption=false,font=footnotesize,labelfont=sf,textfont=sf]{subfig}
\else
  \usepackage[caption=false,font=footnotesize]{subfig}
\fi

\usepackage{url}

\usepackage{comment}
\usepackage{hyperref}
\usepackage{listings}
\usepackage{xcolor}
\definecolor{stringRed}{rgb}{0.64,0.08,0.08}
\definecolor{codegreen}{rgb}{0,0.6,0}
\definecolor{backcolour}{rgb}{0.95,0.95,0.92}

\lstdefinestyle{pythonStyle}{
    commentstyle=\color{codegreen},
    keywordstyle=\color{blue},
    stringstyle=\color{stringRed},
    basicstyle=\ttfamily,
    breakatwhitespace=false,         
    breaklines=true,                 
    captionpos=b,                    
    keepspaces=true,
    showspaces=false,                
    showstringspaces=false,
    showtabs=false,                 
    tabsize=2    
}

\begin{document}

\title{Automated Parallel Kernel Extraction from Dynamic Application Traces}

\author{Richard~Uhrie,~\IEEEmembership{Member,~IEEE},
  Chaitali Chakrabarti,~\IEEEmembership{Fellow,~IEEE}, and
  John Brunahver,~\IEEEmembership{Member,~IEEE}}
\markboth{IEEE Transactions on Parallel and Distributed Systems}{Uhrie \MakeLowercase{\textit{et al.}}: Automated Parallel Kernel Extraction from Dynamic Application Traces}

\IEEEtitleabstractindextext{
  \begin{abstract}

    Modern program runtime is dominated by segments of repeating code called kernels.
    Kernels are accelerated by increasing memory locality, increasing data-parallelism, and exploiting producer-consumer parallelism among kernels - which requires hardware specialized for a particular class of kernels.
    Programming this hardware can be difficult, requiring that the kernels be identified and annotated in the code or translated to a domain-specific language.
    This paper describes a technique to automatically localize parallel kernels from a dynamic application trace, facilitating further code optimization.

    Dynamic trace collection is fast and compact. With optimization, it only incurs a time-dilation of a factor on nine and file-size of one megabyte per second, addressing a significant criticism of this approach.
    Kernel extraction is accurate and performed in linear time within logarithmic memory, detecting a wide range of kernels.
    This approach was validated across 16 libraries, comprised of 10,507 kernels instances. To validate the accuracy of our detected kernels, five test programs were written that spans traditional kernel definitions and were certified to contain all the kernels that were expected.

  \end{abstract}

  }

\maketitle

\IEEEdisplaynontitleabstractindextext

\IEEEpeerreviewmaketitle

\section{Introduction}

\IEEEPARstart{T}{echnology} scaling has slowed and continued increases in application-performance must look for innovation beyond an improved switching device\cite{esmaeilzadeh2012dark}.
Thus, innovation in architecture has focused on deriving higher throughput-performance from the energy-efficiency of specialized hardware\cite{hameed2010understanding}.
This specialization has driven system diversity, with modern Systems on Chip (SoCs) being populated by a heterogeneity of accelerators\cite{kim1988general}.
Consequently, migrating a software-base to such a system is labor intensive, requiring a programmer to map and optimize each stage of an application-pipeline to a specific hardware unit.
In this paper we present a tool that allows us to detect and describe an application's parallel kernels, which would facilitate both the selection-of and mapping-to heterogeneous accelerators.

Applications are composed of a producer-consumer interconnection of kernels\cite{gramps} \cite{asanovic2006landscape} (section \ref{sec:kernelDefinition}).
Parallel kernels are a semantic collection of basic blocks that are clustered temporally, recur many times, and can be executed simultaneously without distorting the result. They are composed of various programmatic structures which include loops, recursion, or library calls. They dominate the execution time of an application. Kernels account for over 99.9\% of all code executed in our test applications (figures \ref{fig:tSize} and \ref{fig:tTime}).

Current approaches for detecting kernels usually require that code be preformatted to make kernels explicit. High Level Synthesis (HLS) relies upon hand annotated code with compiler directives indicating kernel structures. Domain specific languages (DSLs) like CUDA\cite{CUDA} and Halide\cite{halide} function as wrappers that put kernels into a format that labels them explicitly. Static analysis techniques such as polyhedral analysis\cite{polly} detect some kernel types but often struggle to find recurring structures that aren't written as a loop with static boundaries. These limitations mandate that input code for polyhedral analysis be written with static analysis tools in mind. Detecting kernels in naive code is currently not viable for most programmatic structures. Dynamic traces are powerful and can enable better program optimization\cite{wasabi}\cite{ernst2003static} or profile an application to enable introspection of the computation (Dissegna\cite{Dissegna} and Pin\cite{pin}); however, dynamic traces have not been used extensively thus far due to their relatively high expense in trace time, trace size, and analysis time\cite{113085}\cite{zhai2011efficiently}. To fully analyze naive code, a new approach is needed.

\begin{figure}
  \centering
  \includegraphics[width=1\linewidth]{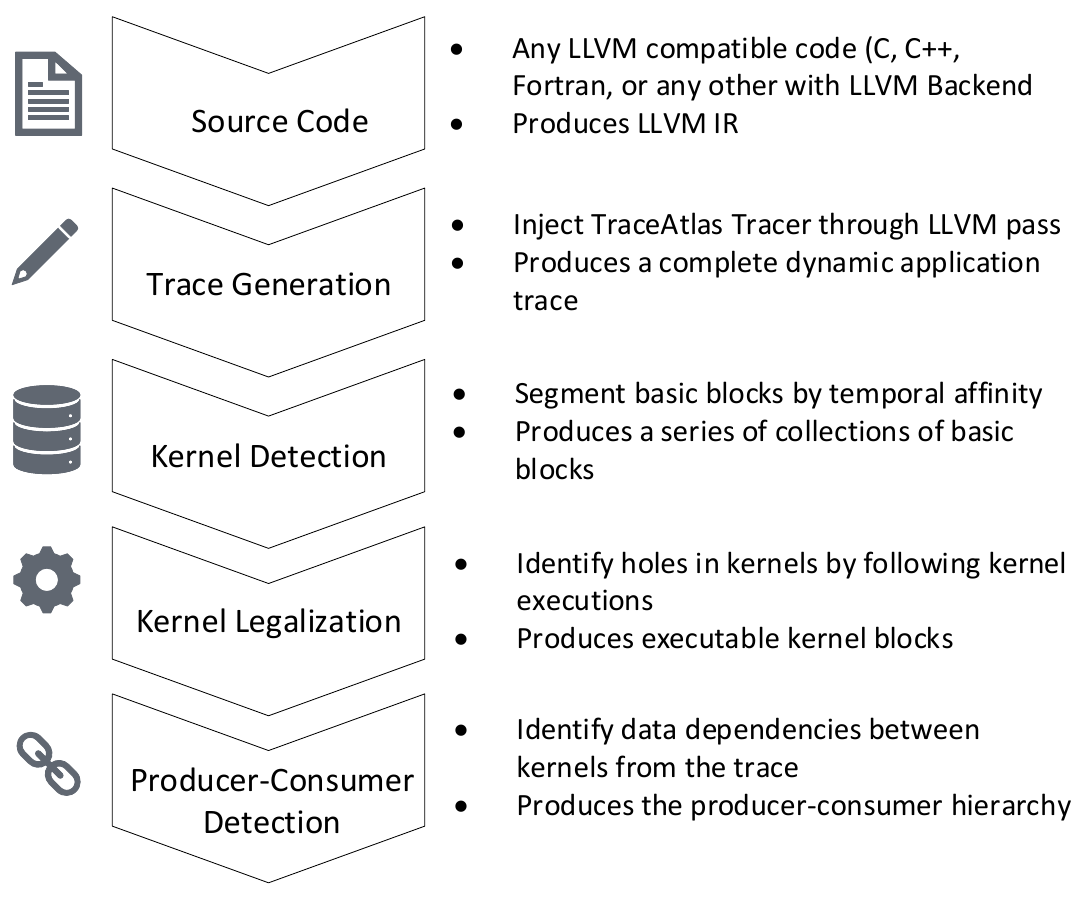}
  \caption{TraceAtlas Analysis Pipeline: The TraceAtlas analysis process is split up into intermediate steps. A tracer is injected into LLVM IR to generate a dynamic trace. From the trace, kernels are detected based on temporal basic block adjacency. Kernels are then legalized before analyzing memory to detect the producer-consumer relationships.}
  \label{fig:overview}
\end{figure}

We have developed TraceAtlas, a tool that enables the tracing of an application with a time dilation factor of only nine. TraceAtlas can also detect kernels from naive, unformatted code with only ten megabytes of memory. The detected kernels can then be analyzed to find the true producer-consumer relationship between kernels based on execution rather than expression. Figure \ref{fig:overview} demonstrates the entire pipeline for TraceAtlas to take source code and label the producer-consumer relationship between kernels. First, LLVM IR is annotated to produce a dynamic application trace. This trace is then analyzed temporally to determine the affinity between basic blocks; the detected collections are then compared to the execution of the trace to produce kernels. Finally, the memory accesses of the kernels within the trace are analyzed to determine their producer-consumer relationships.

We present a method to extract parallel kernels for unannotated LLVM-compatible languages using dynamic trace analysis.
Such a tool would support innovation across many use-cases.
The dynamic trace is created by selectively emitting compressed LLVM-intrinsics at run-time (section \ref{sec:trace}).
We have developed an algorithm to detect the kernels from the input application trace using a log space algorithm that permits reading of the trace in minutes, (section \ref{sec:kExtr}). The results of our trace improvements and kernel extraction are available in section \ref{sec:results}.

We evaluate our approach by analyzing four different design methodologies used to write kernels: For loops, recursion (section \ref{sec:caseLoop}), kernel libraries (section \ref{sec:caseLibrary}), and interface libraries (section \ref{sec:caseComples}). In addition, we compared the results of our analysis of a radio Orthogonal Frequency-division Multiplexing (OFDM) system to the kernels the expert predicted and achieved identical results (section \ref{sec:caseOfdm}).

Our tracing technique has a speedup of 230 times over naive implementation and 49 times over current state of the art. Our algorithm has successfully identified over 10,000 kernels in 392 applications from 16 libraries.
We discuss the limitations of our approach and enumerate some of the potential uses for our tool's dynamic traces and kernels extracted in section \ref{sec:discussion}.

The contributions of this work are:
\begin{itemize}
  \item TraceAtlas\footnote{Available from CodeOcean at \url{https://codeocean.com/capsule/2cb73b4e-11f9-4547-8fe3-4b4956d3d251/} and GitHub at \url{https://github.com/ruhrie/TraceAtlas}}: an open source tool for fast-in-time and small-in-space dynamic tracing of applications
  \item A logarithmic space, linear time algorithm for the extraction and legalization of kernels from a dynamic application trace that can approach terabytes of data
  \item A formal definition for parallel kernels which encapsulates most other kernel definitions while describing some of the important kernel attributes
\end{itemize}

\section{Background and Motivation}

Kernels represent the vast majority of the program computation and are one of the most important code segments to optimize.
A kernel is composed of a semantic collection of basic blocks in a program that are clustered temporally and recur many times.
Currently, kernels are only detectable if the programmer wrote their code in a very specific way that modern tools expect.
Because of this, modern kernel-based tools require kernels to be written explicitly, either through code annotations or using DSLs.
The objective of this work is to demonstrate a method of detecting and extracting kernel code from unformatted code.
Dynamic tracing would allow for simple detection of these kernels, but current techniques are inefficient to the point of failure on larger programs and often cannot trace the entire program.
To dynamically detect kernels in an application, dynamic tracing is a promising technique, but it has received limited use thus far.

The basic blocks that compose a kernel are primarily hot code. By optimizing the hot code, large performance improvements can be made. Greendroid\cite{greendroid} utilized a hot code detector to identify the portions of code which represented large portions of the energy consumption. By identifying and transferring this computation to specialized hardware, they were able to save significant energy. This hot code forms the seed of a kernel in our approach and can be used to identify temporally recurring code segments.

To discover kernels, it is necessary to have a formal definition. Numerous projects have built independent custom DSLs that have a kernel definition specific for their type of kernels. Gramps\cite{gramps} and Halide\cite{halide} focus on image processing kernels. Regent\cite{regent} emphasizes the types of kernels found in high performance computing. RAPID\cite{RAPID} contains kernel structures that are specific to FPGA syntheses. PaRSEC\cite{parsec} specializes in scheduling streaming kernels on heterogenous architectures. StreamIt\cite{streamit} calls every kernel a filter and creates output data by pulling it from the output, requesting each producer in the chain to create the required data. This is just a small subset of DSLs that have been developed to represent kernels. All of these techniques ask the user to write the body of the kernel explicitly as well as some metadata that specifies how they are connected. A holistic kernel definition must bridge these definitions.



The discovery of parallelism in an application is currently focused on the extraction of individual loop-parallelism through code annotation and formatting. Polyhedral analysis tools like LLVM-poly\cite{polly} are able to detect loops in code through static code analysis; however, techniques such as these are limited to loop detection and will require that the loops be written in a way that the tool recognizes. HLS works around this by requiring that all loops of interest be manually annotated. This technique allows the programmer to specify exactly what should happen to their code, with the caveat that the user must know how to optimize the code themselves and what kernels are in their code.

Current static analysis tools fail to detect several important variants of kernels. Dynamic loop boundaries make it impossible to know how parallel a loop is without fully executing it. Recursion is another example where the number of recursions is not inherently obvious and may be difficult if not impossible to determine. An FFT for example will recurs down $\log_2{n}$ times and this behavior cannot be determined without knowing n in advance, something that may or may not be known at compile time.


A dynamic trace is a history of the computation that includes what operations occurred in what order.
Aladdin\cite{aladdin} created dynamic traces from C applications to predict the performance of a custom-built ASIC based upon its C implementation. By analyzing everything the application executed, accurate predictions about the expected cost to perform the computations were able to be made. Wasabi\cite{wasabi} is another dynamic tracing tool that created traces from web assembly code to enable further static analysis techniques. Aladdin is only able to detect single kernels from an application, requiring that the source code be prepared in advance. Aladdin also takes a long time to trace larger applications due to a lack of tracing optimizations. Wasabi improves upon this by tracing all input code and having a time dilation factor of only ~70, but they only support high level WASM, not lower level operations.

Dynamic traces have seen some limited use as an extension to JIT. If kernels are labeled in advance, a minimal dynamic trace that primarily stores address information can be used to detect dependencies\cite{8665818}. This currently relies upon having the kernel labels available in advance, and without this information. it must all be held in memory. This makes JIT analysis infeasible for larger applications and larger kernel optimizations.


Once kernels are known, a myriad of static program optimizations can be accomplished. The simplest optimization is to reformat the kernels into an alternative that is more easily digested by polyhedral analysis\cite{polly}. The extracted kernels can be identified as a target for optimization by genetic algorithms\cite{stochastic}. Similarly, because the kernels represent a large proportion of the computation, they are an optimal target for approximate computing and can achieve the best improvements. Kernels can also be compiled at this point to run on a more efficient, compatible architecture that specializes in kernels like a GPU or TPU. Kernel classification through ML inference\cite{Deniz2016} can also direct compiler optimization\cite{8357388} or the direct exchange of naive code for expertly tuned library calls such as FFTW. Automatically detecting kernels will also enable identifying emergent classes of kernels that have not yet been expertly categorized.

\section{An Overview of Kernel Properties}

\label{sec:kernelDefinition}
Kernels are a semantic collection of basic blocks that are clustered temporally and recur many times. 
Basic blocks represent unconditionally executed code sequences, thus membership in a kernel applies either to all or none of the instructions within. 
Sub-programs (i.e. functions and inner-loops) are composed of multiple basic blocks executed in sequence, and a kernel is represented by their graphical collection. 
Given the sequential execution of these sub-programs, basic blocks that are executed close-in-time are likely to be members of the same kernel. 

Kernels are executed many times in the course of a program; therefore, basic blocks that compose them must also recur many times. 
To be functional they are not composed exclusively of high frequency basic blocks but also low frequency blocks that occur occasionally in the path of the computation.
Boundary conditions and special case control flow will rarely occur but are still part of the kernel. 
This is functionally different from hot-code, which is composed of basic blocks that have been labeled ``hot'' for executing a sufficient number of times.  
Kernels, should they lack loop-loop dependencies between instances, can be scheduled entirely in parallel or a hybrid of parallel and sequential operation.
Most kernels can be rewritten to remove these dependencies making them intrinsically parallel.
Finally, kernels have a semantic meaning, representing a block of computation with specific function and a produce-consume relationship with peer kernels. 

To summarize, a kernel is an amalgamation of several different code properties. Specifically, kernels are composed of basic blocks that:
\begin{itemize}
  \item are grouped to form collections that are semantically related in purpose
  \item have a high probability of being temporally adjacent in execution
  \item are executed many times
\end{itemize}

\begin{figure}
  \centering
  \includegraphics[width=1\linewidth]{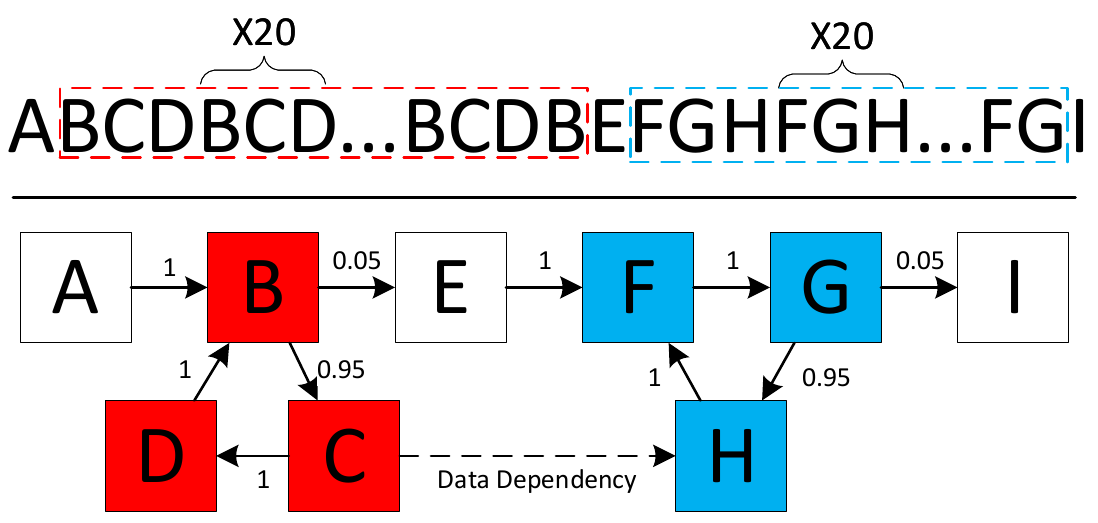}
  \caption{Kernel Basic Block Example: Above is an example temporal ordering of basic blocks in a dynamic trace. Below is the decision-to-decision path (DD-Path) for the computation where each edge weight is the probability of the computation following that path. The first kernel in red is composed of a for loop and repeats twenty times. The second kernel in blue is a recursion that also occurs twenty times and consumes data from the first kernel.}
  \label{fig:kernelBbEx}
\end{figure}

\begin{figure}
  \includegraphics[width=\columnwidth]{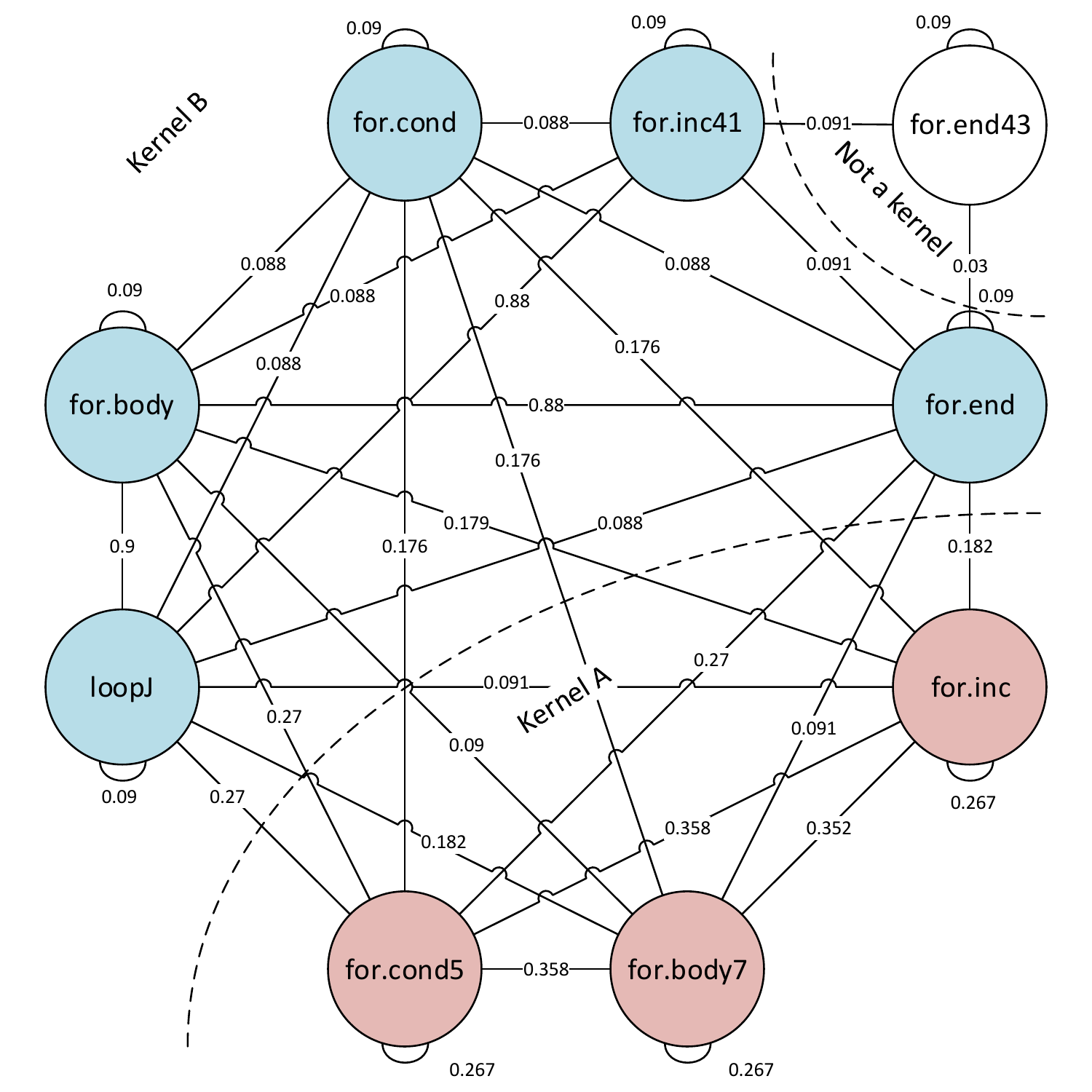}
  \caption{Kernel Probability Graph: The clique affinity graph is built from basic block affinities. Each edge represents the affinity between the basic blocks. This code originated from the MD benchmark in SHOC}
  \label{fig:kClique}
\end{figure}

Our definition captures all parallel kernels defined by Asanovic et al.~\cite{asanovic2009view}. 
For-loops, the prototypical kernel, contain the temporal adjacency between the basic blocks and occur as many times as the loop.
Recursive function calls, an atypical kernel configuration, contain the same temporal adjacency and recur for the recursion depth.
Another kernel configuration, a task scheduler (not included in this work), will execute individual kernels for smaller iteration counts, interleaving the execution of kernels.
As long as the kernels are not consistently executed in the same order, the basic blocks in this design layout still maintain the temporal affinity and recurrence behavior of kernels.
Should they be consistently scheduled in the same order, they will appear as a single fused kernel.
Combinatorial logic will only be detected as a kernel should it be deemed as hot code, so it is either not a kernel by our definition or is contained within a loop of some description.
Finite state machines are composed of two primary components: a state controller and the state actions. These will be detected as two different kernels rather than one if the FSM is composed of hot code.

\begin{figure*}[!t]
  \centering
  \includegraphics[width=1\linewidth]{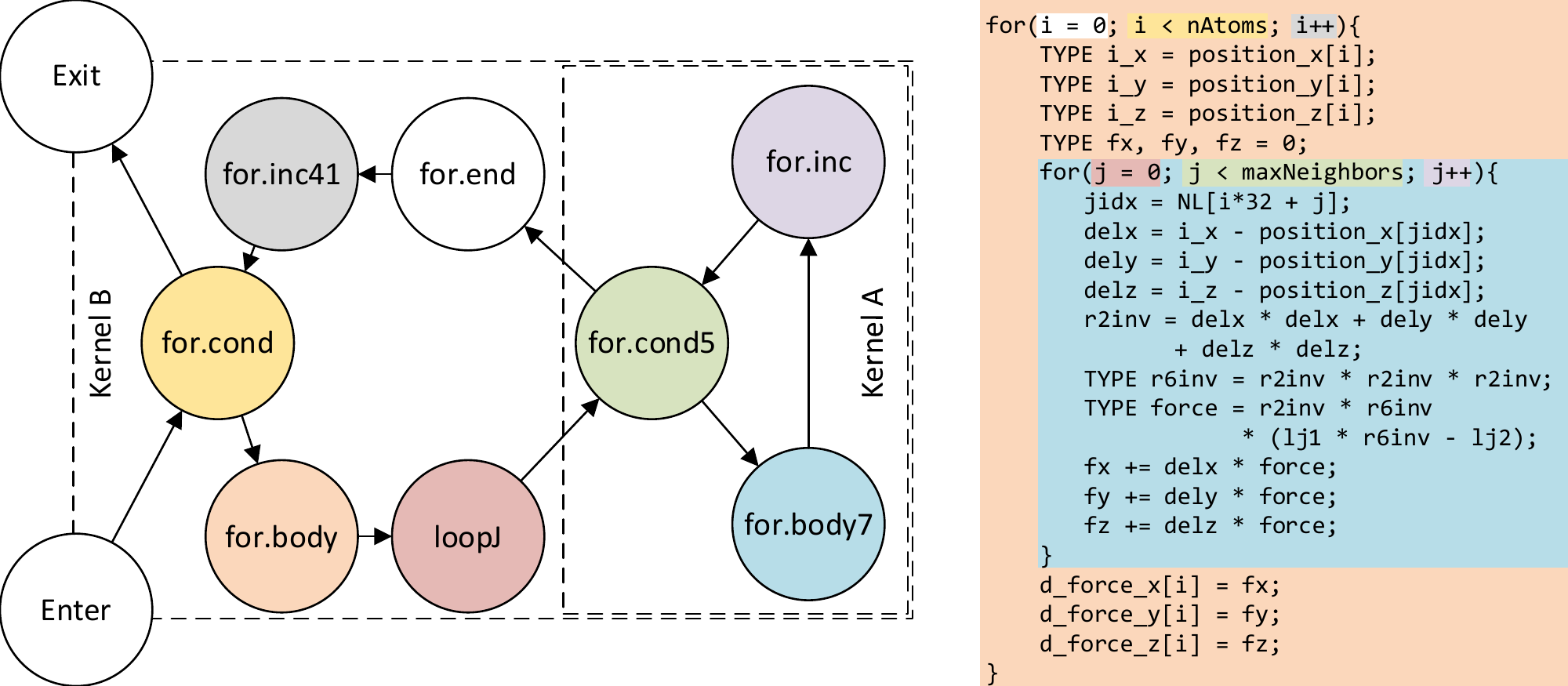}
  \caption{Kernel Basic Block Diagram with Code: The code on the right is a kernel that was detected from the MD benchmark in SHOC. The diagram on the left represents Decision-to-decision path (DD-path). Each individual basic block is highlighted in a matching color to indicate where the basic block derives its code from. Code that is white or a basic block that is white represent a feature that has no analog in the other domain.}
  \label{fig:kernelFlow}
\end{figure*}

Figure \ref{fig:kernelBbEx} contains an example decision-to-decision path (DD-path) which contains two kernels. The red kernel is composed of basic blocks B, C, and D which are repeated many times over. The blue kernel is later in the pipeline and consumes data from the red kernel. It is composed of basic blocks F, G, and H. These two cycles are prototypical kernels and arise from a loop (red kernel) and a recursion (blue kernel).

This definition of kernels allows for one kernel to be contained within another. Since a kernel is a collection of basic blocks that recurs, should there be another recursion within the kernel, such as within a nested for-loop, the inner kernel will be composed of a subset of the basic blocks within the parent. Additionally, this is a one-way relationship. A kernel contains child kernels, but a kernel can be called in multiple contexts and thus have many parents. An example of this would be an FFT. Traditionally an FFT8 contains two FFT4s and a twiddle kernel. FFT4 and the twiddle are sub-kernels nested within FFT8, but the FFT4 can be called from another kernel as well. This creates a hierarchy of kernels from the outside-in with the outermost kernel being responsible for scheduling and memory operations and the inner kernels recurring over many input data sets.

Figure \ref{fig:kernelFlow} contains an example DD-Path for a computation containing two kernels, one nested within another. The code originates from the SHOC benchmark\cite{shoc} and was acquired from Aladdin's github page\cite{aladdin}. Each basic block in the DD-Path is assigned a color which corresponds to the highlighting color on the right. Kernel A specifically only contains the looping part of the inner for loop, $for.cond5$, $for.body7$, and $for.inc$. Single execution parts of the for loop are excluded because they run only once. Kernel B contains all the basic blocks from Kernel A, thus the blocks of Kernel A are guaranteed to be a subset of the blocks in Kernel B.

\section{Low Overhead Dynamic Traces}
\label{sec:trace}

Dynamic tracing with current techniques is prohibitively expensive in both runtime dilation and disk utilization. A naive trace which only stores executed instructions produces a large amount of data, often exceeding 100 gigabytes for a single-second program. A few simple, high-level optimizations dramatically ameliorate this problem. This paper evaluates the following optimizations:

\begin{itemize}
  \item Compressing the trace data with Zlib\cite{aladdin}
  \item Clustering the trace writes so lower os overhead
  \item Encoding trace information before it is compressed with Zlib
\end{itemize}

These optimizations allow us to shrink the execution time to a runtime factor of nine and produce a trace that rarely exceeds five gigabytes. Table \ref{tab:atlasComp} demonstrates the relative improvements and cost of each optimization.

\begin{table}[b]
  \centering
  \caption{Normalized TraceAtlas Performance Comparison}
  \label{tab:atlasComp}
  \begin{tabular}{llll}
    Technique  & Size  & Time   & Effective Change   \\
    \hline
    Naive      & 40.8  & 238.8  & Raw dumping        \\
    Zlib (SoA) & 1.704 & 438.1  & Compress output    \\
    Bursting   & 1.808 & 348.98 & Burst dumps        \\
    TraceAtlas & 1     & 9.1    & Encode information \\
    +Addresses & 7.84  & 248.9  & Add addresses      \\
  \end{tabular}
\end{table}

TraceAtlas generates dynamic traces by injecting logging statements into the LLVM IR. The naive solution is to simply export the IR as the application executes. This generates a trace in plain text and is relatively robust; however, applications today run billions of operations a second. By exporting all information with no processing, the disk always becomes the bottleneck in both write speed and storage size.

Trace data can be compressed as it is generated by Zlib. This is the current state of the art solution used by Aladdin\cite{aladdin}. Trace information is extremely low in information and can thus achieve exceptionally high rates of data compression. This optimization moves the bottleneck from disk write speed to processor execution. Due to the high repetitiveness of the data, Zlib was able to easily identify the vast majority of the repetition in the traces. Different compression levels did not appear to appreciably impact trace size or time. Overall, this optimization was found to double the execution time, but shrink the trace size by ~20x on small applications. Larger applications dramatically increase the runtime due to a Zlib operating better on smaller kernels. We do not have numbers on this due to the inability of naive tracing to produce a trace that fit on the hard drive, but intermediate applications had an overhead ranging from 500-2000x relative to the original execution.

Exporting information to the trace as soon as it becomes available is relatively expensive due to the system call overhead. There are two primary overheads in dumping the trace information: the call overhead to export the information itself and compressing the data with Zlib to write it to disk. Since basic blocks always run in order, it is possible to only dump all basic block trace data at the end of the trace. This removes many potential export calls that scale with the size of the basic block. Similarly, Zlib compresses most efficiently when it has a relatively large chunk of data to process. By waiting until there are 4-128 kb of data available lowers the cost of using Zlib to compress the trace. The combined effect of these two optimizations resulted in a 25\% trace time improvement.

Ultimately, Zlib compresses the trace such that the basic blocks are being encoded. This information is known at compile time, and if we do this manually, we can get significant performance improvements. By assigning a key to every basic block and exporting that information instead of the actual IR, the compression effort is significantly reduced. This efficiently encodes the trace before it is given to Zlib while maintaining critical information on the structure of the computation. The result was that the trace size halved again while the total time to trace fell to an overall time dilation of nine.

Depending on the demands of your application, additional information may be required beyond the path of the execution that add some additional overhead. For our purposes, we were interested in looking at the memory dependencies between kernels in order to extract the producer-consumer relationships between kernels, see section \ref{sec:discussion}. In order to do this, we also exported the addresses of all loads and stores. This resulted in the trace time rising dramatically and a significantly larger trace due to Zlib having more difficulty compressing the information due to how addresses vary. Aladdin's implementation exports the addresses of all load and store operations, but an in-house variant was written with support for additional features.  As a result, addresses are not represented in the numbers reported in Table \ref{tab:atlasComp}\cite{aladdin}. Additional trace values can also be exported for an additional overhead.

TraceAtlas has minimal overheads and can trace every application with LLVM IR. Table \ref{tab:atlasComp} contains an overview of our performance improvements with the average trace size and average time dilation factor for all applications that were able to be traced naively with less than a terabyte of disk and traced with Zlib in under 48 hours. The cumulative effect of all our optimizations resulted in a runtime speedup of 400x on small programs versus SoA and over 100,000x for larger programs. Trace sizes halved for smaller applications and 1500x for large programs. These levels of overhead put execution time of an application to nearly real time. As a result, traces can be generated quickly with minimal data stored. This alleviates the costs of dynamic tracing and makes it a viable tool to perform additional optimizations, such as the identification of kernels.

\section{Kernel Extraction}
\label{sec:kExtr}
A kernel is a subsection of a program that executes many times throughout the lifespan of a program. This subsection gets transformed into basic blocks as part of the compilation process. As a result, a kernel can be represented as a collection of basic blocks that are connected in some type of looping structure. This may include a standard loop, a recursive function, some other cyclic structure in their DD-Path, or a combination of these three.

Detecting these kernels is a multi-step process. The first step is to cluster the basic blocks from the source application into a series of temporally adjacent basic blocks using a greedy, heuristic algorithm (section \ref{sec:kernelDetect}). Once segmented, these collections will potentially contain duplicate kernels or be missing individual basic blocks. To fix this a smoothing algorithm must be applied to remove these errors and transform them into kernels (section \ref{sec:smoothKernel}).


\subsection{Kernel Detection}
\label{sec:kernelDetect}

The defining attribute of kernels is that there is a high degree of temporal affinity between basic blocks within the kernel. If a single basic block from a kernel within a trace is examined, there is a high probability that the prior basic blocks and subsequent basic blocks will also be a part of the kernel. We refer to this as the basic block affinity which can be most easily calculated as $P(A\alpha ^r B^k)$. This is the probability that basic block $B$ occurs $k$ times within a distance $r$ of $A$. Distance refers to the number of basic blocks that are executed between the current basic block and the target basic block.

More generally, basic block affinity is the probability that any one basic block occurs within a range of another. Our approach uses a uniform weighting over an execution window $r$ of seven basic blocks, but other PDFs can be used to achieve similar results. A window refers to the range of basic blocks that are analyzed during the computation. Our results in section \ref{sec:results} informed our choice of window width.

Using this affinity metric, we can calculate a score between any two basic blocks by summing over the size of the window as is done in equation \ref{eq:affinity}. The final score is subsequently divided by the size of the window to make the final scores mimic a probability. A kernel will then be defined as a summing of basic blocks such that the mutual sum of the score of all basic blocks in a set are greater than a certain threshold probability.

\begin{equation}
  f_r(A,B)=\frac{1}{2r+1}\sum_{k=1}^{2r+1}P(A\alpha ^r B^k)
  \label{eq:affinity}
\end{equation}

This approach which is summarized by equation \ref{eq:affinity} is advantageous because it only requires storing the previous $2r+1$ basic block IDs in memory. Due to the potential of traces to contain billions of basic blocks or more, a logarithmic space algorithm is mandatory to analyze the trace. This approach scales linear with the number of basic blocks in the source application and logarithmically with the size of the trace. Since the trace size is dramatically larger than the number of basic blocks in a typical application, equation \ref{eq:affinity} satisfies this constraint.

With the affinity scores calculated, one can then represent the collection of basic blocks within a program as a fully connected digraph with edge weights equal to the given score. Figure \ref{fig:kClique} contains an example set of values where each edge is the maximum of the two edges between the nodes. Within Figure \ref{fig:kClique}, each kernel is composed of a set of basic blocks such that the sum of the weight from every node within a kernel to every other node within a kernel sums to at least the target threshold, 0.95 in this scenario. 

The structure of this affinity allows us to only store the ID of the last $2r+1$ basic blocks. By using this small amount of memory, we can increment counters for each basic block within a range $r$ the basic block at memory location $r$. This generates the affinity numbers, $P(A\alpha ^r B^k)$. This algorithm only utilizes logarithmic space and linear time, allowing for quick analysis of traces that can exceed terabytes of raw data.

Figure \ref{list:decodeDetect} presents a method for calculating the sum given in equation \ref{eq:affinity} with minimal memory. As a trace is streamed through, commit the prior $2r+1$ blocks to memory. A counter in a matrix is then incremented at row $priorBlocks[r]$ and every value in $priorBlocks$ including $r$. Finally each value is normalized by the number of times that block occurs.

\begin{figure}[ht]
  \begin{lstlisting}[language=Python]
for block in trace:
  for i in 1 to (2r + 1):
    bCount[priorBlocks[r]][priorBlocks[i]]++
    priorBlocks.shiftBack(block)
  counter[block]++
for block in bCount:
  for other in bCount:
    prob[block][other] = \
      bCount[block][other] / counter[block]
    \end{lstlisting}
  \caption{Basic Block Affinity Calculation: This code efficiently reads in a trace and calculates the affinity calculation from equation \ref{eq:affinity} for every basic block in the trace. It has been tested on traces over a terabyte in size and executes in under a day.}
  \label{list:decodeDetect}
\end{figure}

With the affinity score between basic blocks calculated, we can segment the graph by utilizing a greedy algorithm as given in Figure \ref{list:decodeDetect2}. To do this, the algorithm iterates through basic blocks, using them as a seed to create kernels. We first sort the basic blocks by frequency count in decreasing order. Then for every basic block that hasn't already been added to a kernel is used as a seed of a new kernel, and we greedily add the basic block with the highest affinity score to the new kernel until we reach a desired threshold, currently 0.95. The resultant collection of basic blocks will be such that there is at least a 95\% chance that you will see another block from the kernel for every instance of a block in the kernel. Upon reaching the threshold each of the blocks within the kernel gets added to a set of explained blocks which are no longer valid seeds for new kernels. Finally, once the count is under the hot code threshold (512 in this case) the algorithm terminates. Formal graph cuts were found to not be necessary to properly segment the computation.

\begin{figure}[ht]
  \begin{lstlisting}[language=Python]
explainedBlocks = set()
blockCount.sort()
for block, count in blockCount:
  if count < 512:
    break
  if block in explainedBlocks:
    continue
  sortedRow = prob[block].sort()
  kernel = []
  score = 0
  while score < 0.95:
    kernel.add(nextBlock)
    score = GetScore(kernel)
  for a in kernel:
    explainedBlocks.add(a)
    \end{lstlisting}
  \caption{Basic Block Fusion: With basic blocks affinities calculated, each basic block above the hot code threshold needs to be clustered to form kernels, 512 in this example. This listing greedily adds basic blocks to a kernel until the mutual affinity of all basic blocks in the kernel is greater than the given threshold, 0.95 in this example. }
  \label{list:decodeDetect2}
\end{figure}

To get more consistent results, it is advantageous to sort basic blocks in decreasing order. Nested kernels are guaranteed to have a higher execution count than that of the parent kernels. By sorting the seed basic blocks by execution count, we guarantee that we will find kernels from the inside out and will not skip detecting a nested kernel because we identified the parent first.

It is necessary to exclude certain basic blocks from being seeds because the affinity scores can result in a distinct set of blocks that do not form a kernel, specifically in nested kernels. If a basic block from a parent kernel is selected as a seed and it is on the interface between the child and parent kernels, it is possible for the affinity to be slightly higher with the child than with the parent. This will result in a new kernel that is the child kernel with this basic block on the interface included. It is distinct from both the parent and child kernels and cannot be removed with the algorithm in section \ref{sec:smoothKernel}. Fortunately, not every basic block needs to be used as a seed, but rather only enough blocks to explain all the hot code. The basic block collections that are produced from this algorithm have the potential to have holes in them or even detect duplicate kernels. To resolve this, a second stage algorithm must be used to refine the kernels into a more useful form.

\subsection{Kernel Legalization} 
\label{sec:smoothKernel}


The basic block collections detected in section \ref{sec:kernelDetect} are only an abstract set of basic blocks that have a high probability of being temporally adjacent. This definition, although it takes advantage of some of the features of kernels, is mathematical and fails to properly represent the functional aspect of the source code.
There are two primary flaws in the detected kernels: only one path of a conditional may appear, and larger kernels may be represented as multiple kernels. Both occur because the prior algorithm was only looking for temporal affinity. Kernels must be able to execute in a continuous path, which is not guaranteed based on temporal affinity alone.

\begin{figure}
  \centering
  \subfloat[Conditional Code]{\includegraphics[width=\columnwidth]{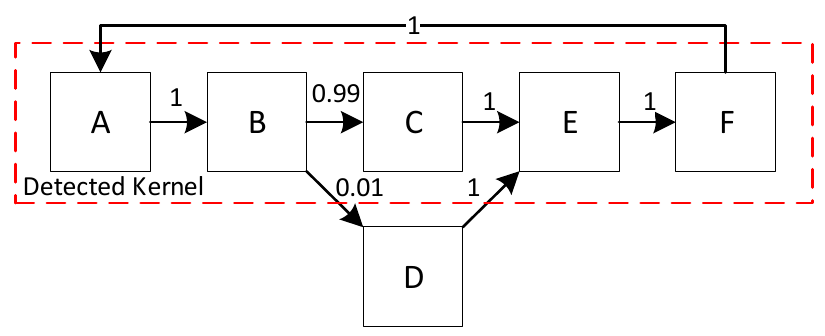}\label{fig:conditionalFlaw}}\\
  \subfloat[Long Kernel]{\includegraphics[width=\columnwidth]{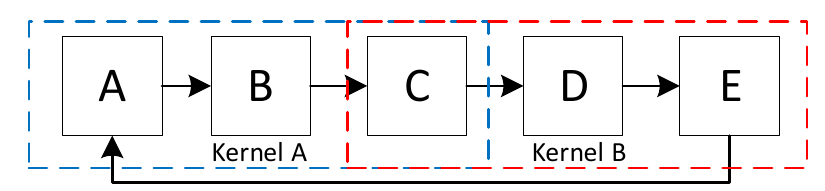}\label{fig:widthFlaw}}
  \caption{Kernel Detection Flaws: The kernels detected at this point have two flaws which can be corrected. \ref{fig:conditionalFlaw} is an example DD-Path of edge case conditional logic where the rare branch is excluded due to its probability being too low. \ref{fig:widthFlaw} contains two separate kernels where the algorithmic threshold was reached before absorbing every basic block in the kernel, allowing a second kernel to be detected based on the remnants.}
  \label{fig:kernelIssues}
\end{figure}

If a kernel contains a conditional block, these blocks will usually be detected; however, if one of the paths happens less often than $1-threshold$ there is a possibility that it will not be selected by the greedy algorithm. This is prone to happen relatively often in image processing kernels as the edges of the image often require a conditional to either zero pad the data or extend the size of the image. Figure \ref{fig:conditionalFlaw} contains an example DD-Path for just such an example where there is a conditional branch that occurs 1\% of the time and is thusly not selected as part of the kernel.

Kernels that are significantly wider that the algorithm window will often be detected multiple times due to the nature of seed selection. The set of basic blocks representing this kernel may overlap minimally or not at all. Figure \ref{fig:widthFlaw} is an exaggerated example of this occurring with an analysis radius of one for a five-block wide kernel. The first kernel selects basic block B as its seed and hits the threshold using just basic blocks A, B, and C. Basic block D is unrepresented, so it then gets chosen as a seed and hits the required threshold be using blocks C, D, and E. Both kernels are valid detections, but they both represent the same kernel and need to be fused.

To legalize a kernel, we need to find all the basic blocks that a kernel may potentially enter between iterations without entering a different kernel. The smoothed kernel can be collected by following the trace once more now that we know the kernel sets. Our kernel legalization algorithm streams through the trace and creates a set of basic blocks for each kernel that contain the basic blocks we have seen since we were last in the kernel block set. If we re-enter the it before entering another kernel, we know these blocks are a part of the kernel, so we add these blocks to the set. If we enter a new kernel, we know we have exited the current kernel and should clear its contents.

This algorithm adds every basic block that occurs within the body of a kernel without adding basic blocks from other kernels. Nested kernels are still detected successfully with this algorithm because the set of new blocks are cleared when we encounter the beginning of a kernel again. Kernels that contain more than double the algorithm window will result in at least two identical sets of basic blocks which can be trivially detected as identical. Conditionals within the middle of a kernel are fused directly. Conditionals from the end of a kernel must ultimately reenter the kernel before a new iteration is started to branch back to the beginning of the kernel due to the way LLVM schedules conditionals.

\section{Results}
\label{sec:results}
TraceAtlas is an open source tool that can successfully dynamically trace an application with the primary limitation being the runtime of the program. We have an average runtime dilation factor of nine and produce one megabyte for every second of execution. The resultant traces can be efficiently analyzed to detect kernels using a greedy algorithm to identify the flavor of kernels of interest to the user. Over 7000 kernels were identified from 392 programs originating from 13 open source projects.

\subsection{Methodology}

All of the applications traced for this paper were compiled with clang 8.0 and linked with lld 8.0. Zlib 1.2.11 was used for compression. Python 3.6 was used for the kernel detection algorithms. Table \ref{tab:kernels} enumerates the collection of libraries and benchmarks used for tracing, the applications using that library, the number of kernels detected, and the version of the library or benchmark. The libraries were selected to sample a variety of general kernel-based tools.

The applications were run on Xeon E5-2650 processors with the trace data being written to an Intel SSD DC S3500 with 1TB of storage. Each application was executed nine times and the median value was reported to filter out noise. Each sample was given 48 hours to execute and was allowed to consume an entire terabyte of storage. The application samples that exceeded these limits were canceled and are not present in Table \ref{tab:atlasComp} or Figures \ref{fig:tSize} and \ref{fig:tTime}.

\begin{table}
  \centering
  \caption{TraceAtlas Test Corpus}
  \label{tab:kernels}
  \begin{tabular}{llll}
    Library                                     & Applications & Kernels & Version \\
    \hline
    Gnu Scientific\cite{galassi2002gnu}         & 15           & 705     & 2.6     \\
    FFTW\cite{frigo2005fftw}                    & 3            & 63      & 3.3.8   \\
    Eigen\cite{eigenweb}                        & 16           & 7823    & 3.3.6   \\
    OpenCV\cite{bradski2008learning}            & 21           & 1516    & 4.1.0   \\
    LiquidSDR\cite{gaeddertliquid}              & 88           & 1288    & 1.3.1   \\
    FFmpeg\cite{bm12}                           & 2            & 221     & 4.2     \\
    Perfect Benchmark\cite{perfect}             & 13           & 93                \\
    SHOC Benchmark\cite{shoc}                   & 10           & 25                \\
    Armadillo\cite{sanderson2016armadillo}      & 24           & 207               \\
    StreamIt Benchmark\cite{thies2010empirical} & 11           & 165               \\
    GraphBLAS\cite{bulucc2017design}            & 1            & 24      & 3.1.1   \\
    mbed TLS\cite{l15}                          & 8            & 881     & 2.16.3  \\
    FEC\cite{karn2007fec}                       & 8            & 239     & 3.0.1   \\
    Dhrystone and Whetstone\cite{DAW}           & 3            & 183               \\
    Cortex Suite\cite{Thomas2014CortexSuiteAS}  & 21           & 422               \\
    spuce\cite{spuce}                           & 33           & 171     & 0.4.3   \\
    Total                                       & 392          & 10507             \\
  \end{tabular}
\end{table}

\subsection{Motivating Use: Producer-Consumer Pipeline Extraction}

With kernels fully discovered, the same trace as before can be used to analyze real memory dependencies between kernels. By tracking loads and stores, it is possible to identify which kernel instances wrote to and read from a particular address.

A producer-consumer pair occurs when a producer stores data that the consumer reads from. Specifically, the producer creates a store that puts a value into memory at a specific address. This address is now most recently written to by that kernel instance and any kernel instance that loads data from that address is a consumer of that instance.

Detecting the memory dependencies between kernels can also be done with a small memory overhead that is bounded by the memory used during the original computation. By going through the trace, it is possible to note when you enter and exit a kernel, thus denoting individual kernel instances. Then, for every store one can write an instance ID to a dictionary where the address is the key. Every load can then check if this address is in the dictionary and if it is then the current kernel instance knows it read from the matching value in the dictionary.

A limited top-level producer-consumer pipeline extractor tool has been developed and is available with TraceAtlas. This was used to aid in the verification of the kernels detected by TraceAtlas and to generate many of the figures in this work.

There are many uses of kernels with a producer-consumer graph that are not explored in this work. With the producer-consumer graph, kernels can be fed as an input into static analysis tools like LLVM-Poly\cite{polly} to raise their performance. Furthermore, a compiler can be written to transform the detected kernels into code compatible with other kernel libraries to perform fusing in TACO\cite{kjolstad2017tensor} or scheduling in Halide\cite{halide}. In short, the kernels that this tool detects can be easily fed into other tools to perform additional analysis or further raise the performance with no expert programmer involvement.

\subsection{Dynamic Tracing}


%
Our techniques resulted in a reduction of trace size by a factor of fifty relative to naive techniques and a reduction of trace time by a factor of two relative to Zlib compression. Each strategy approached a different problem and achieved a speedup in its domain and often for others as well. The net improvement of all these optimizations has resulted in a tracing technique that runs within a reasonable time dilation.

Two variants of the TraceAtlas tool have been developed: one that traces the path of an execution and another that also traces the memory address of every load and store. This is the only technique in Table \ref{tab:atlasComp} that encodes address information. Each subsequent optimization also enables additional applications to be traced due to lower performance limitations. Clustered basic block dumping allowed for the tracing of some shorter cpp programs in Eigen. The current version of TraceAtlas has no currently known limitations beyond only supporting single threaded applications and the performance overhead of tracing.

\begin{figure}[ht]
  \includegraphics[width=\columnwidth]{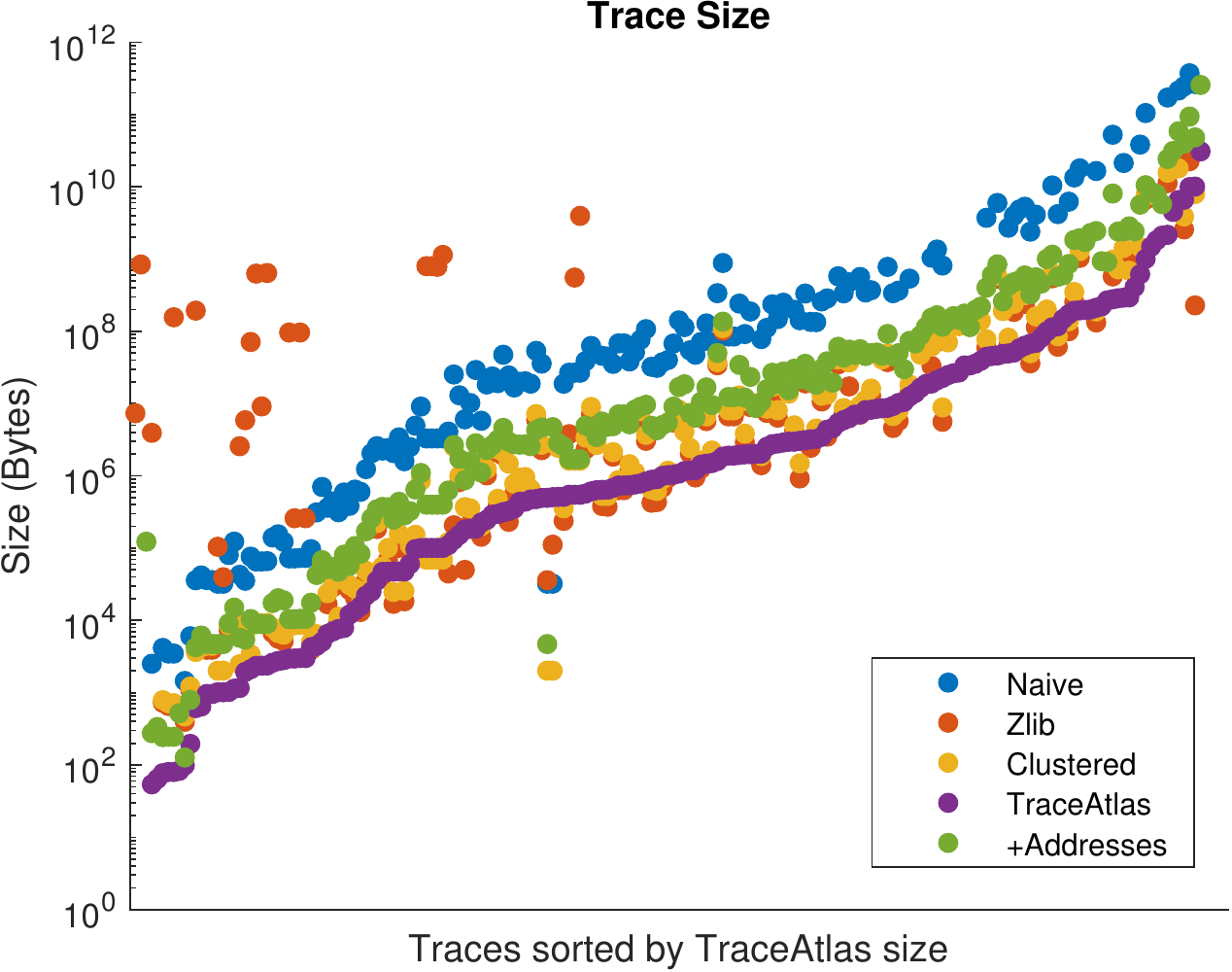}
  \caption{Overall Trace Size: The horizontal axis sorts traces by the size of the TraceAtlas trace. Usually, TraceAtlas is significantly smaller than other techniques. On occasion other techniques will perform better due to the implementation of Zlib, but on average it produces traces that are significantly smaller.}
  \label{fig:tSize}
\end{figure}

Figure \ref{fig:tSize} shows that TraceAtlas, as a rule, produces smaller traces due to the information being compressed statically before being fed to Zlib. Occasionally, raw Zlib compression and IO clustering will produce a smaller trace, likely due to the method Zlib uses to compress the trace being more efficient that our current encoding scheme. For additional details, see section \ref{sec:discussion}. Table \ref{tab:atlasComp} shows that on average, TraceAtlas produces a trace that is half the size of what is produced from raw Zlib compression.

\begin{figure}[ht]
  \includegraphics[width=\columnwidth]{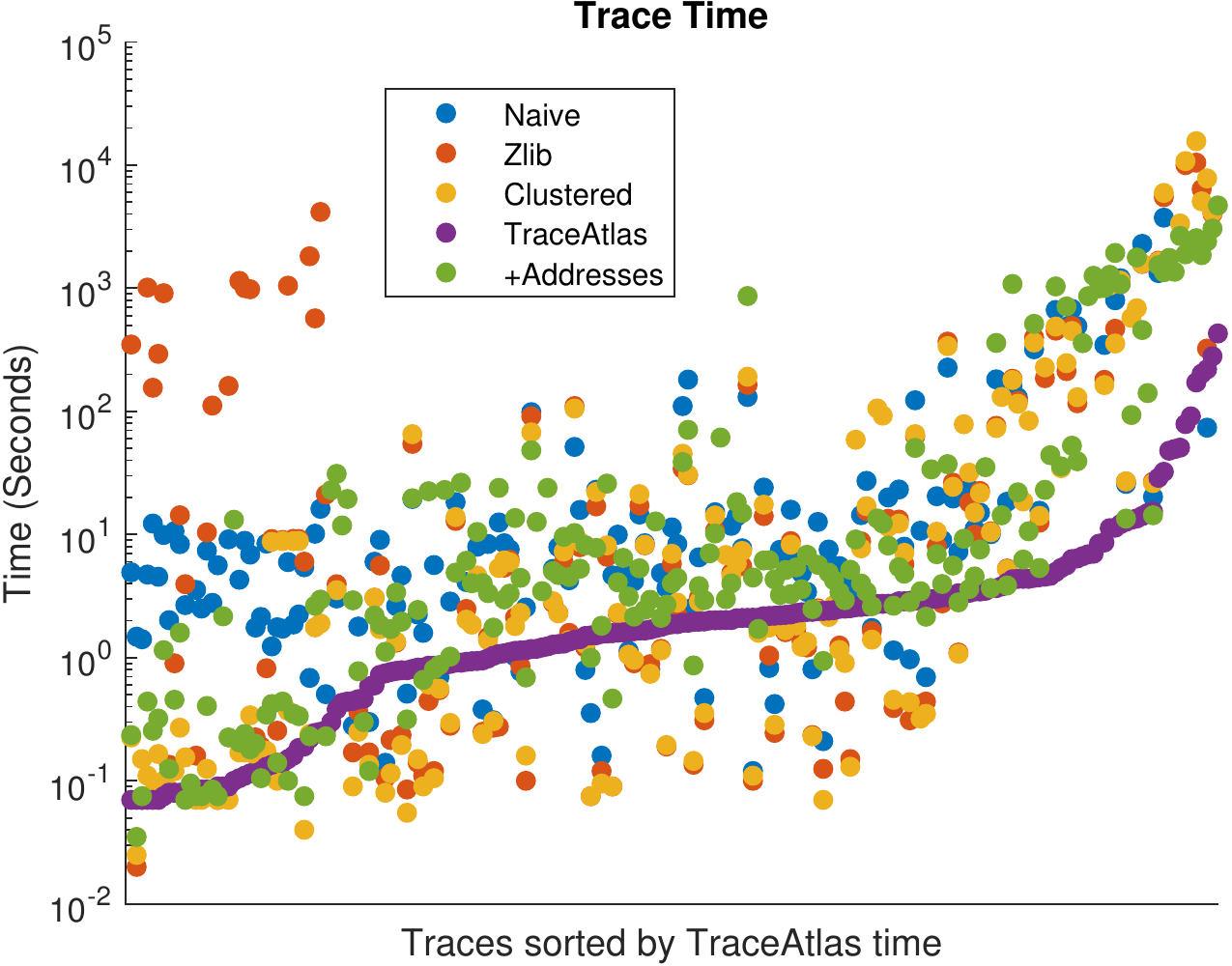}
  \caption{Overall Trace Speed: The horizontal axis sorts traces by the speed of the TraceAtlas trace generation. Usually, TraceAtlas is significantly faster than other techniques. On occasion other techniques will perform better due to the implementation of Zlib, but on average it runs faster. Larger applications, however, perform dramatically worse than TraceAtlas.}
  \label{fig:tTime}
\end{figure}

Figure \ref{fig:tTime} shows that TraceAtlas performs above average across the space; however, once it takes more than a second to trace, TraceAtlas performs significantly faster. When the trace takes TraceAtlas more than a second, it had a time dilation factor of 10.18 while Zlib compression had a factor of 265.5. This shows that TraceAtlas is twenty-six times faster than Zlib compression for larger applications. For additional analysis see section \ref{sec:discussion}.

\subsection{Kernel Extraction Algorithm}

\begin{figure}
  \includegraphics[width=\columnwidth]{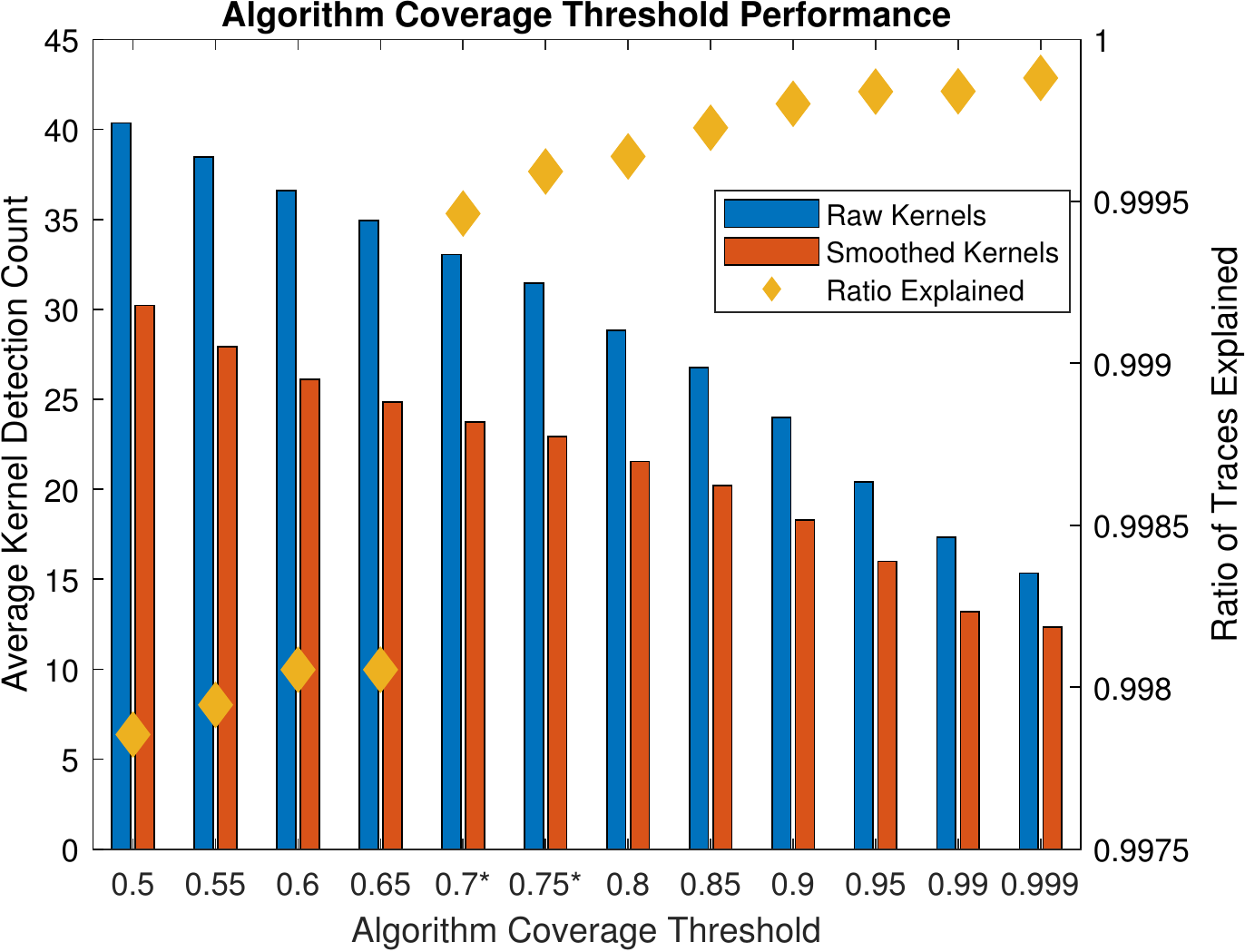}
  \caption{Threshold Parameter Effect: The bars refer to the average number of distinct kernels detected in a trace with respect to the algorithm coverage threshold. The ratio of code explained jumps between 0.65 and 0.7. This indicates that 0.7 or 0.75, marked with an asterisk, are optimal values to detect the most succinct kernels  while explaining the majority of the code.}
  \label{fig:kThresh}
\end{figure}

\begin{figure}
  \includegraphics[width=\columnwidth]{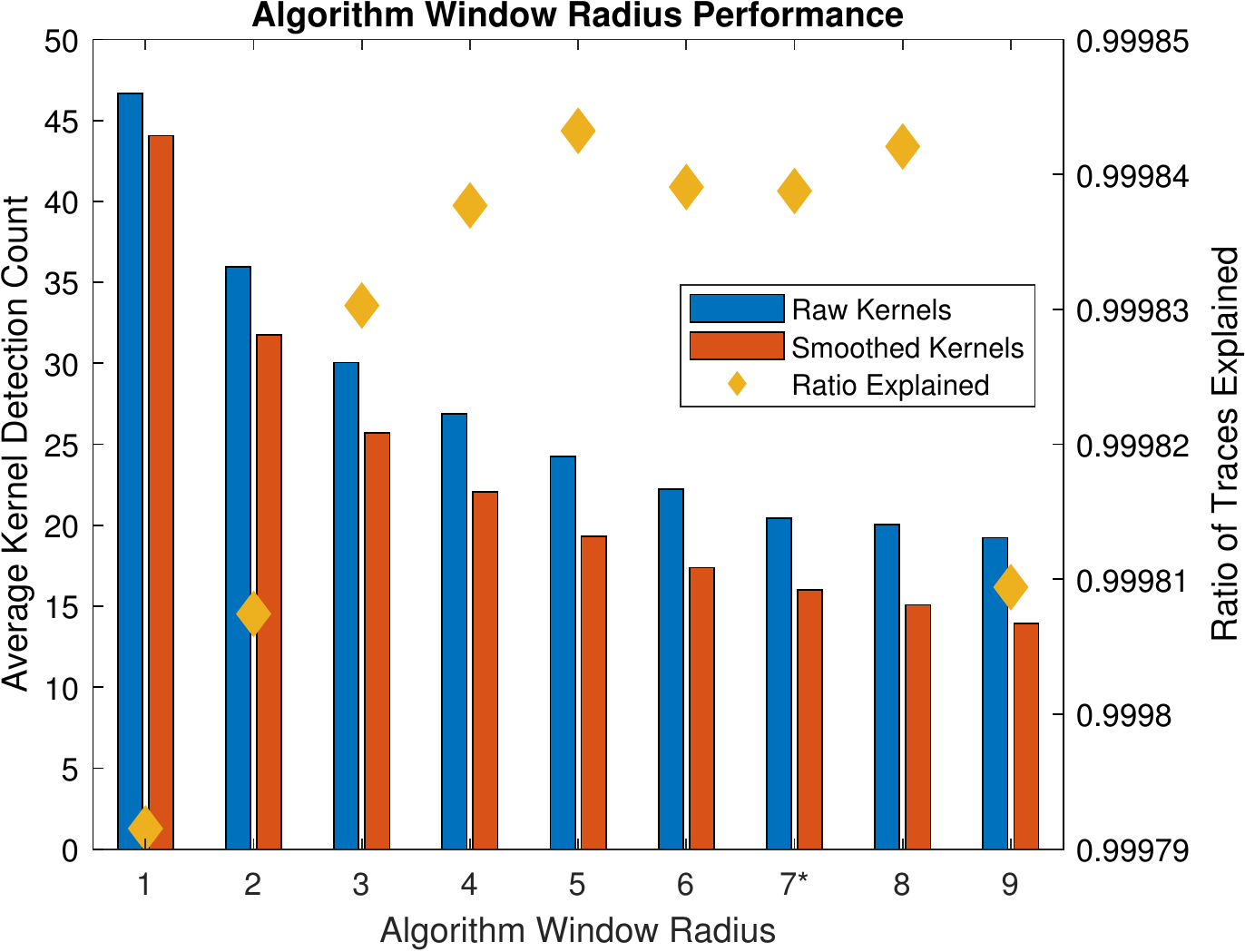}
  \caption{Window Radius Parameter Effect: The bars refer to the average number of distinct kernels detected in a trace with respect to the algorithm window radius. The ratio of code explained plateaus at a radius of 6 and 7, marked with an asterisk. Due to the lower ratio surrounding this region, a radius of 6 or 7 are recommended.}
  \label{fig:kWin}
\end{figure}

\begin{figure}
  \includegraphics[width=\columnwidth]{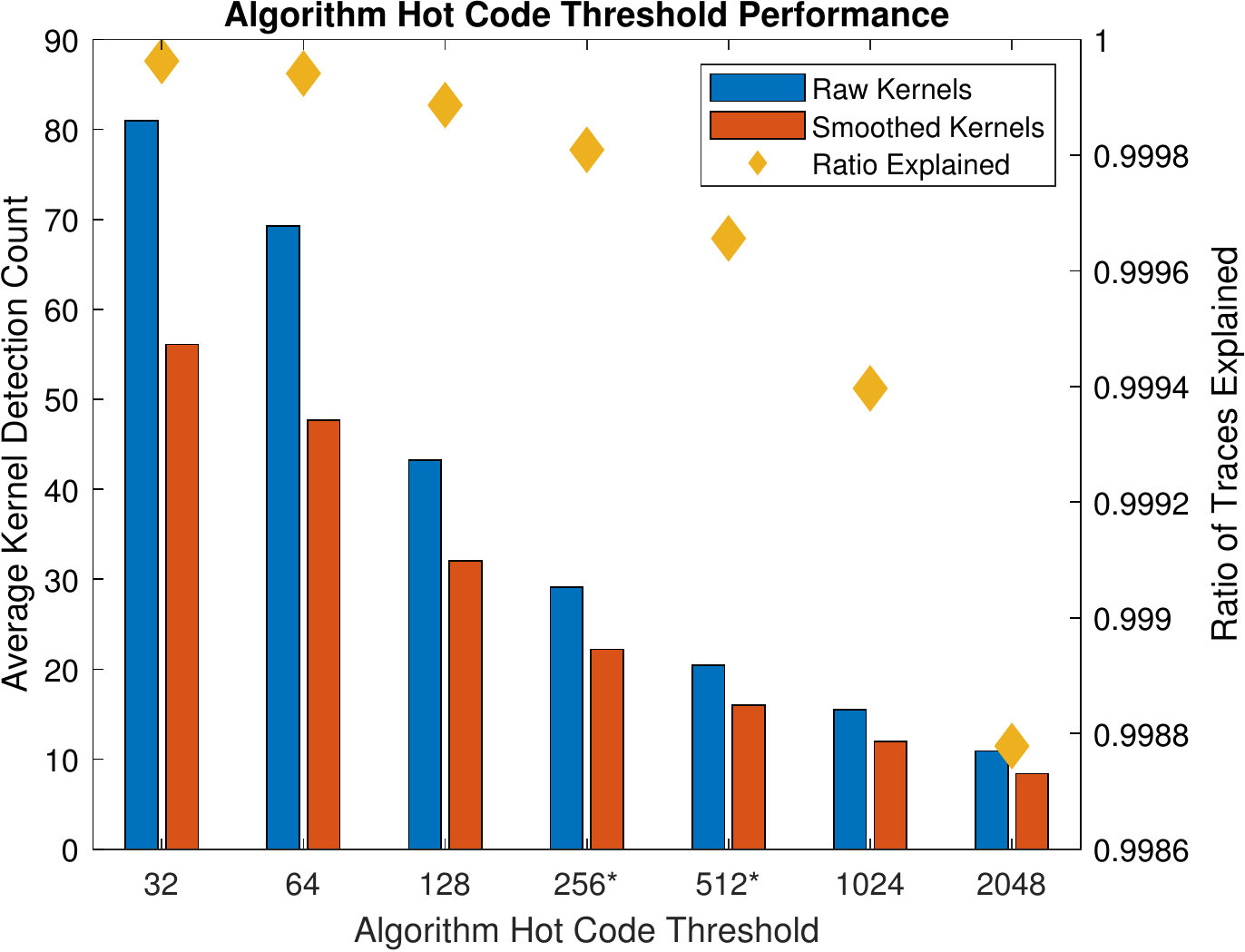}
  \caption{Hot Code Threshold Parameter Effect: The bars refer to the average number of distinct kernels detected in a trace with respect to the hot code threshold. This is a tuning parameter which should be selected based upon the types of kernels the user wishes to detect. A threshold of 256 or 512, marked with an asterisk, fuses most trivial kernels together while still retaining most of the radio of traces explained. }
  \label{fig:kHot}
\end{figure}

The greedy algorithm described in section \ref{sec:kernelDetect} has three tuning parameters: the window radius, the probabilistic sum threshold, and the minimum iteration count to be deemed hot code. In the following graphs, these values were set at a radius of 7, a threshold of 0.95, and a hot code threshold of 512 iterations unless otherwise specified. The ``Ratio of Traces Explained'' in Figures \ref{fig:kThresh}, \ref{fig:kWin}, and\ref{fig:kHot} refers to the ratio of the basic blocks within traces that are contained within our kernels divided by the total number of basic blocks in every trace summed together.

Ideally, we would like to extract as many kernels as possible while simultaneously explaining as much of the program as possible;however, different values for these parameters will cause kernels to be fused, lowering the kernel count while increasing the explained trace ratio. Depending on the specific application and use case, different values maybe required. The optimal values found for our corpus are marked with an asterisk in Figures \ref{fig:kThresh}, \ref{fig:kWin}, and \ref{fig:kHot} for ease of identification.

The algorithmic threshold refers to the minimum likelihood for the greedy algorithm to terminate. Figure \ref{fig:kThresh} shows that the number of raw and smoothed kernels detected decreases in an approximately linear fashion, while the Trace Ratio experiences a significant jump at a probability of $0.7$. Because of this we deem $0.7$ to be the minimum recommended value for this parameter with $0.75$ also being reasonable. Higher values will continue to improve the desired explained trace ratio, but the number of kernels will continue to fuse, potentially obscuring kernels that may be of interest.

The window radius in Figure \ref{fig:kWin} refers to the maximum distance for a basic block to be away from another for them to be deemed adjacent. Figure \ref{fig:kWin} demonstrates that there is a rapid fusion of kernels as the window radius grows from one to four before leveling out. There is a similar rapid growth in the ratio of the basic blocks explained in this same area, but there is also a trailing off effect once the window grows to a width of nine. Due to this behavior, the optimal window width lies somewhere between 5-8 with them all achieving similar performance.

The hot code threshold refers to the minimum number of times a piece of code must happen to be deemed a kernel. This parameter is more a matter of user choice where the code behavior informs the decision. Larger applications running across large data sets will achieve more succinct kernels at higher thresholds, but they will also potentially explain less of the overall application. Figure \ref{fig:kHot} shows that lower ratios do explain more of the program with a higher number of kernels, but depending on the nature of the code these kernels have the potential to not be kernels of interest to the user. Given these curves we find a threshold of 256-512 to be optimal.

\section{Case Studies}
\label{sec:caseStudies}

To validate our approach, four categories that are used to write or call kernels in modern code were identified: loop based grammars, recursive function calls, kernel libraries such as FFTW, and pipeline libraries such as OpenCV. Loop based grammars are the most basic types of kernels and are most easily represented as a simple for-loop or while-loop. Recursive function calls are another common technique for custom kernels. If a user is trying to call a common kernel such as an FFT, they are often wrapped in a library to simplify the interface. This can be done on a low level such as in FFTW, or at a higher level such that there is a single function call that calls an entire pipeline of kernels such as in OpenCV.

Upon analysis of these four cases, TraceAtlas successfully identified all the kernels expected in our source applications. It also identified other code sections as kernels which were not predicted but emerged as an artifact of the implementation of the kernels in the source code. Upon further analysis, these artifacts are kernels but were not anticipated by the programmer due to abstractions. As a final verification, a real-world radio application was analyzed, and the detected kernels were compared with those predicted by the author.

\subsection{For Loops and Recursion}
\label{sec:caseLoop}
For loops are the prototypical kernel. It contains two primary components which we needed to detect: the loop body and the loop iterator.

\begin{figure}
  \includegraphics[width=\columnwidth]{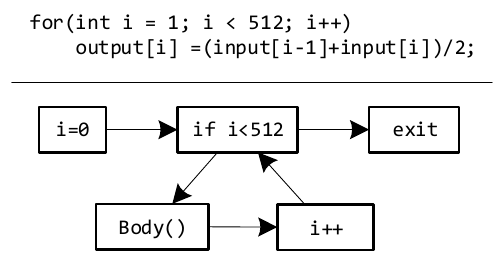}
  \caption{Example For Loop Kernel: The above source code was compiled, executed, and analyzed to detect the kernels within. The LLVM IR contained four blocks which are labeled in the DD-Path below. The kernel which was detected was composed of the cycle in the middle with the initializer excluded because it only happened once.}
  \label{fig:forDag}
\end{figure}

Figure \ref{fig:forDag} contains an example kernel application which averages two adjacent values. Above is the source code and below is the resultant DD-Path. It iterates over 511 points and executes the kernel upon each. When transformed into LLVM IR, a for loop is transformed into four to five basic blocks: one for the initializer, one for the conditional, one for the body, one for the incrementor, and possibly one for the exit.

The portion of this graph that composes our kernel is the cycle present in Figure \ref{fig:forDag}. This cycle is formed from the incrementor, conditional, and body. Upon tracing the code, a kernel is generated where the three basic blocks that compose the cycle are identified as the kernel.

\begin{figure}
  \includegraphics[width=\columnwidth]{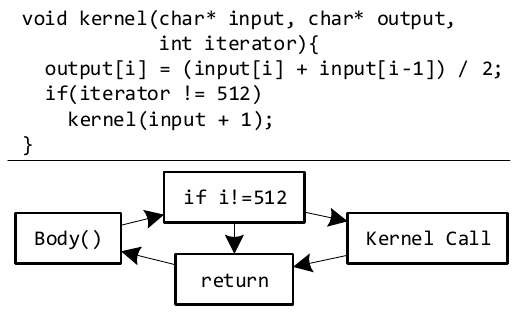}
  \caption{Example Recursive Code Kernel: The above source code was compiled, executed, and analyzed to detect the kernels within. The LLVM IR contained four blocks which are labeled in the DD-Path below. The kernel which was detected was composed of all the basic blocks drawn due to them all happening every iteration. The initialization logic is not written as part of the function call and was not drawn.}
  \label{fig:recursiveDag}
\end{figure}

Recursive functions operate in a similar fashion. They utilize identical attributes: a conditional, a body, and an enumerator which functions as the exit condition. Figure \ref{fig:recursiveDag} contains a kernel that is functionally identical to Figure \ref{fig:forDag}.

Figure \ref{fig:recursiveDag} shows the modified DD-path of the program. The kernel first enters the body and then exits if the exit condition is true ($i == 512$ in this case), otherwise it calls the next kernel instance before exiting.

\subsection{Library}
\label{sec:caseLibrary}
Code written within a library has the potential to be written like that in Figures \ref{fig:forDag} and \ref{fig:recursiveDag}; however, once a user is using a library it is usually to accomplish something either complex or optimized for performance. Both significantly complicate the graph and obfuscate what the kernels are. Fortunately, dynamic traces still extract all the basic blocks that execute though so the repetitive nature of kernels is still easily identifiable.

Our example is a 512-point 1D-FFT scheduled and executed using FFTW based upon code from their website. Upon kernel extraction we identified seven kernels whose blocks are available in Table \ref{tab:fftw}. Upon back referencing the basic blocks to the LLVM IR, one can identify the primary function of each kernel. Kernel 0 reads in input data for the FFT. Kernels 1 and 2 malloc memory for the working set and the output respectively. Kernels 3 and 4 move data into the buffers for the appropriate preparation. Kernels 5 and 6 are responsible for doing the body of the FFT and contain the most blocks. This matches what one would expect for an FFT with additional memory scheduling occurring for performance.

\begin{table}
  \centering
  \caption{FFTW Kernel Blocks Identified}
  \label{tab:fftw}
  \begin{tabular}{l l}
    Kernel Index & Basic Blocks Composing the Kernel              \\
    0            & 4618-4625                                      \\
    1            & 906-912,998,999,1095-1104,1307-1308,1428,1434  \\
    2            & 31,998,1003-1005,4617-4633                     \\
    3            & 390,1003-1022,1459,1461                        \\
    4            & 1012-1017                                      \\
    5            & 1-72,104,158-162,211-257,526-596,1000-1350,... \\
    6            & 1-72,104,158-162,211-257,526-596,1000-1350,... \\
  \end{tabular}
\end{table}

\subsection{Complex application}
\label{sec:caseComples}
More complex libraries further abstract away the concept of a kernel down to a single function call which schedules and executes an entire pipeline of kernels. Brisk\cite{brisk} from OpenCV is an example of this.

\begin{figure}[ht]
  \centering
  \includegraphics[width=0.75\columnwidth]{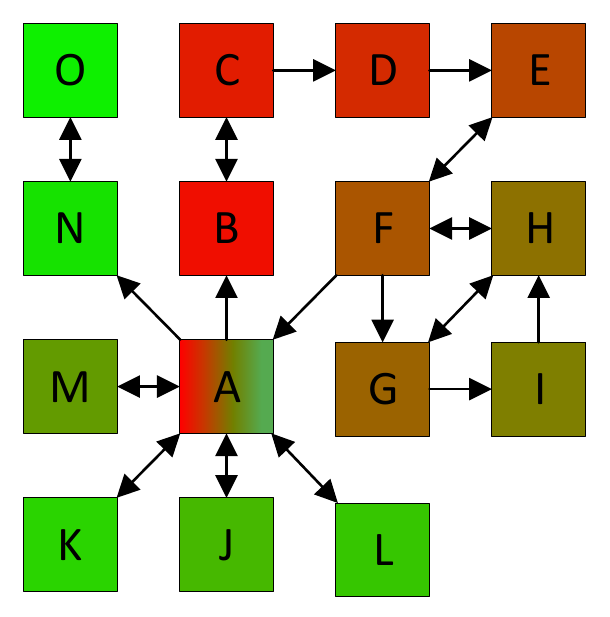}
  \caption{Brisk Kernel Pipeline: Brisk was executed using OpenCV and the top-level producer-consumer pipeline was extracted. Each arrow refers to the ordering of the kernels. Red kernels are executed earlier and green kernels are executed later. Within the pipeline, many kernels hand control to each other frequently. Upon source code analysis, it was found that the Brisk kernel was contained entirely within kernel N with the other kernels responsible for IO, decoding, and memory operations.}
  \label{fig:opencvDag}
\end{figure}

Figure \ref{fig:opencvDag} contains a subset of the kernels detected in the basic Brisk implementation with the kernels that are dominant in the execution present. Red kernels execute earlier in the computation, and green kernels execute later. The primary body of Brisk as described in the paper happens almost entirely within kernel 15 with other kernels being used for scheduling, file IO, and buffer management. There is a single centralized kernel which schedules the other kernels to execute from.

The final extraction contained 33 individual kernels spanning 158,000 basic blocks. This shows that although libraries such as OpenCV perform a lot of computational backend to schedule a kernel, TraceAtlas is still able to detect the kernels that occur and their temporal ordering.

To verify that the graph in Figure \ref{fig:opencvDag} is correct, each of the top-level kernels in the resultant producer-consumer graph were manually analyzed. The parent functions were identified to map them back to the original source. Kernel A is a control kernel which schedules other kernels to execute. Kernels B, C, and D were kernel generation code. E, F, G, H, and I are all responsible for reading in a png file and converting it to accessible buffers. J, K, and L are all responsible for moving memory. Kernel M schedules the kernel to iterate across the color channels, and N performs the actual Brisk computation. Finally, O writes the resultant data to disk.

\subsection{Expert Verified Application: OFDM}
\label{sec:caseOfdm}
\begin{figure}[ht]
  \includegraphics[width=\columnwidth]{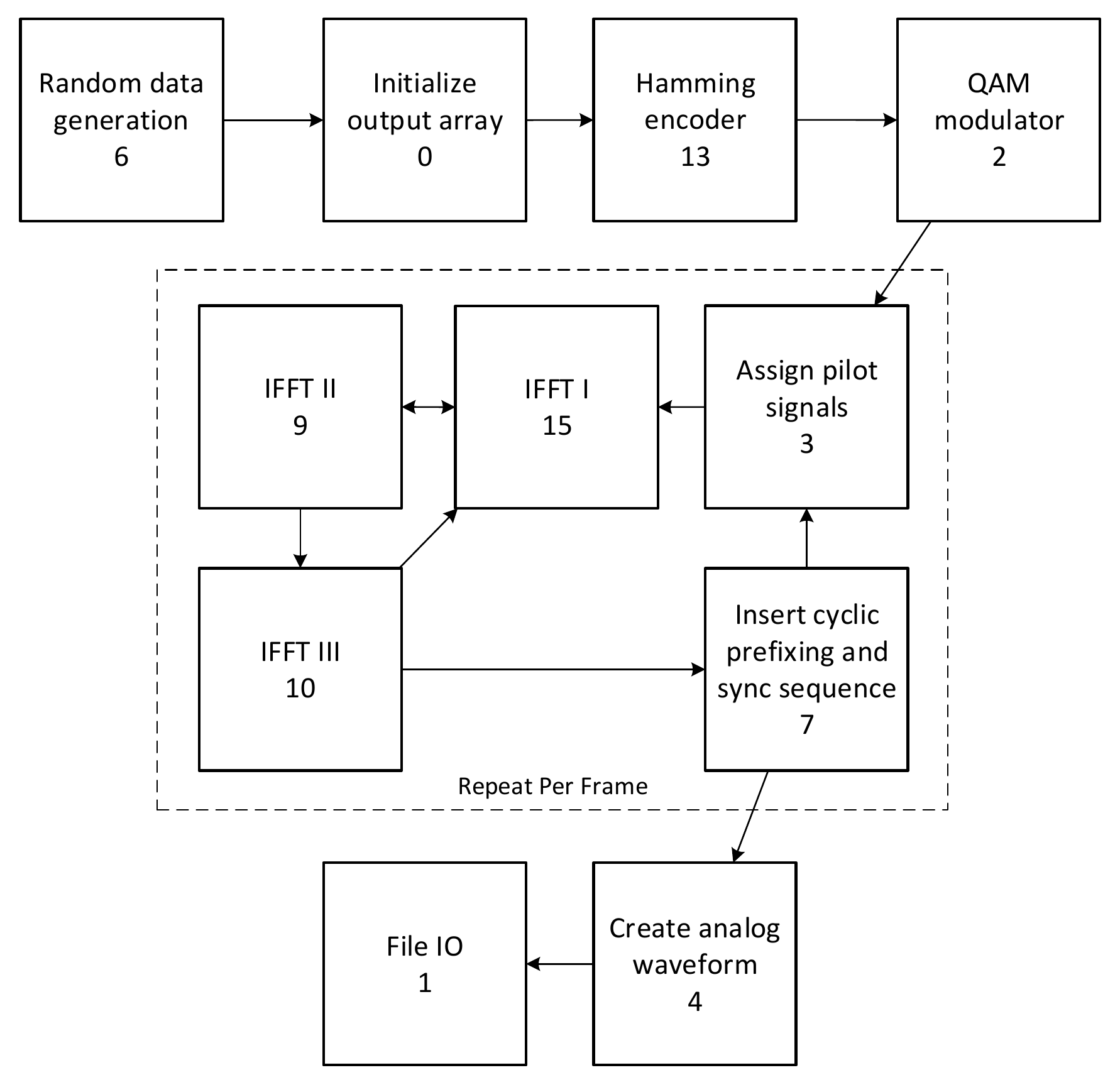}
  \caption{OFDM Kernel Pipeline: This OFDM example program was written externally and analyzed by TraceAtlas. The original author was able to look at the kernels and the producer-consumer relationship and accurately label each of the kernels TraceAtlas detected.}
  \label{fig:ofdmDag}
\end{figure}

As a final verification, an OFDM radio system was acquired. The input application was analyzed with TraceAtlas and a top-level producer-consumer pipeline was extracted from the input application with the detected kernels. The pipeline resulted in Figure \ref{fig:ofdmDag}.

The result was shown to the original programmer who is unfamiliar with TraceAtlas, and they were able to successfully label all the top-level kernels that were detected. This shows that all kernels expected by an expert user are properly represented. Additional nested kernels were detected which were not expected by the user but were found to originate from loops that were deemed insignificant even though they were a large portion of the computation.

This shows that code from the wild can be easily analyzed by TraceAtlas. It successfully identifies kernels in both our input synthetic applications as well as wild code that was not written with TraceAtlas in mind.

\section{Discussion}
\label{sec:discussion}

TraceAtlas has a dependency upon LLVM IR being available for the source application. This limits the use of the tool to C and C++ projects with the potential for Fortran through the use of f18\cite{flang-compiler_2019} or fc\cite{compiler-tree-technologies_2019} or other languages with a custom LLVM frontend. Supporting other interpretive\footnote{Interpretive here refers to languages that require an external binary to execute. Compiled languages with library dependencies such as C\# and Java do not apply.} languages such as Python and Perl are possible if the control binary and supporting libraries are compiled with TraceAtlas injected. Closed source tools can still be used through an IR lifter such as LLVM-mctoll \cite{mctoll}.

The traces produced by TraceAtlas are dynamic application traces. As a result, they will only represent the paths taken in the execution. Unexplored kernels and dead code will not be traced or identified as a kernel in the resultant trace. Care must be taken to ensure that the sample application executes the code of interest and that it executes a sufficient number of times given a user's hot code threshold. Similarly, dynamic traces still require a significant amount of space and time to analyze; however, it is now viable to trace an application.

Asanovic et al. identified many archetypes of kernels\cite{asanovic2009view}. Of the thirteen archetypes generated, our kernel definition (section \ref{sec:kernelDefinition}) successfully identifies twelve of them.
For kernel analysis, one of the preferred attributes is that it composes a significant portion of the computation. In combinatorial logic, it is either contained within a loop and detected, or it is executed sparingly in which case it does not belong in the same category as kernels.
We specifically cannot identify finite state machines. A finite state machine is composed of two different kernels: a state controller and a state handler. Both kernels are detected and properly represented by TraceAtlas, but a finite state machine is not detected as a kernel itself. Multi-threading parallelism through task queues are believed to be compatible with this framework, but the tool chain lacks thread support, so this aspect is yet unexplored.

When analyzing the TraceAtlas performance, some applications achieved better performance with plain Zlib compression than were achieved with our tool. TraceAtlas creates an output trace as a UTF-8 encoded text file where each line is a key-value pair for the point of interest, traditionally the basic block ID. For some applications, it is plausible that the raw Zlib trace is more easily identified than that of TraceAtlas's encoded trace. On average, encoding is advantageous and becomes mandatory for larger traces. This variability within Zlib results in some traces being smaller or even being produced more quickly than is accomplished by TraceAtlas. This could potentially be fixed by changing the encoding of the trace to a more succinct binary format such that Zlib more easily detects repetition within the trace.

\section{Conclusion}

We have demonstrated that dynamic traces are easily generated from application sources. This was tested against 16 libraries and 10,507 individual kernels instances Tracing code can be injected efficiently, and tracing only inserts a time dilation factor of nine. The resultant trace only produced one megabyte of data per second of trace. All of this makes potential dynamic trace tools an inexpensive and plausible solution to many problems.

We have also demonstrated a log space algorithm for analyzing the trace that can detect kernels from within the source application. This technique detects all the input kernels by our definition. We have applied it successfully to hundreds of input applications. We have also demonstrated the resultant data generated from the most common kernel execution types.

These two techniques combined allow for easy analysis and extraction of kernels from the source code based upon its dynamic execution. The resultant kernels can then be extracted and optimized with external tools to achieve even better performance improvements with no user interaction.

\section*{Acknowledgment}
\addcontentsline{toc}{section}{Acknowledgment}

This material is based on research sponsored by Air Force Research Laboratory (AFRL)
and Defense Advanced Research Projects Agency (DARPA) under agreement number
FA8650-18-2-7860. The U.S. Government is authorized to reproduce and distribute
reprints for Governmental purposes notwithstanding any copyright notation thereon.
The views and conclusion contained herein are those of the authors and should not be
interpreted as necessarily representing the social policies or endorsements, either
expressed or implied, of Air Force Research Laboratory (AFRL) and Defence Advanced
Research Projects Agency (DARPA) or the U.S. Government.

The author would like to thank Seth Abraham, Umit Ogras, and Carole Wu for helping edit this paper. The author would also like to thank Liangliang Chang, Mukul Gupta, Vamsi Lanka, Sriharsha Uppu, and Benjamin Willis for helping write input applications for the application corpus.

\bibliographystyle{ieeetr}
\bibliography{IEEEabrv,sources}

\begin{IEEEbiography}[{\includegraphics[width=1in,height=1.25in,clip,keepaspectratio]{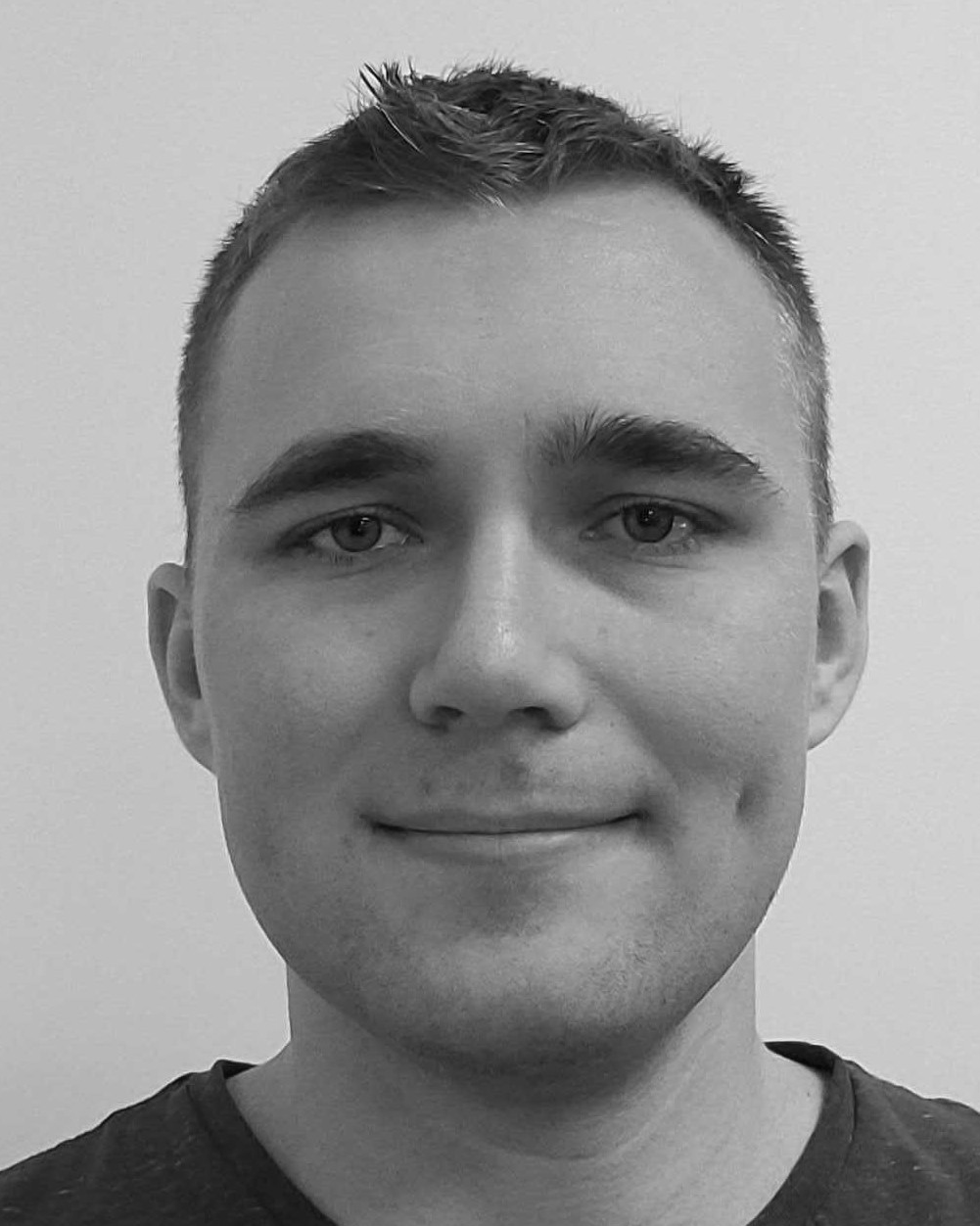}}]
  {Richard Uhrie} received the BS degree in electrical engineering from the Colorado School of Mines in 2016 and the MS degree in computer engineering from Arizona State University in 2018. He is currently a PhD candidate in computer engineering at Arizona State University. His dissertation topic is on detecting and characterizing the structure and behavior of kernels.
\end{IEEEbiography}
\begin{IEEEbiography}[{\includegraphics[width=1in,height=1.25in,clip,keepaspectratio]{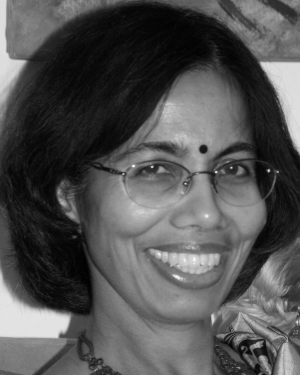}}]
  {Chaitali Chakrabarti} received the BTech degree in Electronics and Electrical Communication Engineering from the Indian Institute of Technology, Kharagpur, India, in 1984, and the PhD degree in electrical engineering from the University of Maryland, College Park, in 1990.
  She is a professor with the School of Electrical, Computer and Energy Engineering, Arizona State University, Tempe.
  Her research interests include VLSI algorithm-architecture co-design of signal processing and communication systems and all aspects of low-power embedded systems design. She is a Fellow of the IEEE.
\end{IEEEbiography}
\begin{IEEEbiography}[{\includegraphics[width=1in,height=1.25in,clip,keepaspectratio]{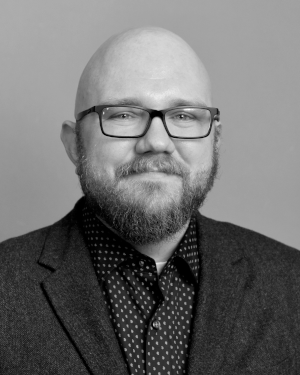}}]
  {John Brunhaver} is an Assistant Professor at Arizona State University in the School of Electrical Computer Energy Engineering as of 2015.  His research focuses on developing a machine understanding of computation, VLSI-design productivity, and radiation-hardened circuits and architectures.  His Stanford University Ph.D. thesis, The Design and Optimization of A Stencil Engine, examines the virtual machine model for an image processing and image understanding domain-specific processor.
\end{IEEEbiography}
\end{document}